\documentstyle[prd,aps]{revtex}
\begin{document}
\preprint{LANCS-TH/9412, SUSSEX-AST 94/11-1, NI94026, astro-ph/9411104}
\draft

\title{False vacuum inflation with a quartic potential}
\author{David Roberts}
\address{School of Physics and Materials, University of Lancaster, Lancaster
LA1 4YB,~~~U.~K.}
\author{Andrew R.~Liddle}
\address{Astronomy Centre, School of Mathematical and Physical Sciences,
University of Sussex, Falmer, Brighton BN1 9QH,~~~U.~K.\\
and\\
Isaac Newton Institute, 20 Clarkson Road, Cambridge CB3 0EH,~~~U.~K.}
\author{David H.~Lyth}
\address{School of Physics and Materials, University of Lancaster, Lancaster
LA1 4YB,~~~U.~K.\\
and\\
Isaac Newton Institute, 20 Clarkson Road, Cambridge CB3 0EH,~~~U.~K.}
\date{\today}
\maketitle
\begin{abstract}
We consider a variant of Hybrid Inflation, where inflation is driven by two
interacting scalar fields, one of which has a `Mexican hat' potential and the
other a quartic potential. Given the appropriate initial conditions one of
the fields can be trapped in a false vacuum state, supported by couplings
to the other
field. The energy of this vacuum can be used to drive inflation, which ends
when the vacuum decays to one of its true minima. Depending on parameters,
it is possible for inflation to proceed via two separate epochs, with the
potential temporarily steepening sufficiently to suspend inflation. We use
numerical simulations to analyse the possibilities, and emphasise the
shortcomings of the slow-roll approximation for analysing
this scenario. We also
calculate the density perturbations produced, which can have a spectral index
greater than one.
\end{abstract}
\pacs{PACS numbers: 98.80.Cq, 04.50.+h\\ \vspace*{10pt} \noindent
\hspace*{2cm} LANCS-TH/9412, SUSSEX-AST 94/11-1, NI94026, astro-ph/9411104}

\section{Introduction}

Recent models of the inflationary cosmology have increasingly adopted two
scalar fields, utilising the increased complexity to develop a more varied
phenomenology than traditional chaotic inflation models based on a single
field \cite{CHAOTIC}. Where one of the fields is coupled explicitly to the
Ricci scalar, one has the `Extended Inflation' model \cite{LaSteinhardt} and
its generalisations \cite{extinf}, which were introduced in order to
revive the idea of a trapped scalar field providing the energy density to
drive inflation. However, it is also perfectly possible to realise this
notion without leaving Einstein gravity, by taking advantage of couplings
between the two scalars \cite{Linde90,AdamsFreese}.  For suitably chosen
potentials, the interaction term can hold one scalar field trapped in a false
vacuum state while the other evolves; eventually this evolution triggers a
phase transition, which typically ends inflation within a Hubble time
and leads to rapid reheating.

The simplest potential realising  this `hybrid  inflation' scenario
is\footnote{In general $\psi$ can be multi-component with full
angular symmetry, in which case its appearance in this paper should be
interpreted as $|\psi|$.}
\begin{equation}
V(\phi,\psi) = \frac{\lambda}{4} (\psi^{2} - M^{2})^{2} +
\frac{\lambda'}{2} \phi^{2} \psi^{2} + \frac12 m^2 \phi^2
\,.
\end{equation}
After its proposal by Linde \cite{LIN2SC}, this potential was investigated by
several authors \cite{LL2,LIN2SC2,MML,CLLSW}. The fact that it is quadratic
in the inflaton  field, together  with the existence of the false vacuum,
means that it can emerge naturally in the demanding context of superstring
motivated  supergravity \cite{CLLSW}. Very few models of inflation have
this distinction\cite{ewan}.

In this paper we consider the potential
\begin{equation}
V(\phi,\psi) = \frac{\lambda}{4} (\psi^{2} - M^{2})^{2} +
\frac{\lambda'}{2} \phi^{2} \psi^{2} + \frac{\lambda''}{4} \phi^{4} \,.
\end{equation}
It differs from the earlier one in that it is quartic instead of quadratic,
in the inflaton field $\phi$. Though this potential does not have any clear
particle physics motivation, we feel it to be worth investigating because it
gives an inflation scenario with several novel features.

We have not investigated the possibility that both a quartic and a quadratic
term play an important role, because that would occur only in a special
region of the already rather special permitted region of parameter space. A
recent treatment of this more general parameter space has been carried out by
Wang \cite{WANG}, which surveyed the possibilities for generating unusual
density perturbation spectra. However, in that paper a detailed investigation
of the models was not carried out; for instance the possibility of inflation
proceeding in two separate epochs was not discussed. Further, as we shall
show the traditional slow-roll analysis is not sufficient to outline the full
range of possibilities, and some regions which possess slow-roll inflationary
solutions are not accessible when one considers more general evolution.
Finally in this connection we note that the particular form of potential we
study is not covered by Wang's treatment as she utilised field redefinitions
which are singular when the quadratic term is absent.

\section{The two field potential}

The absolute minima of the fields exist at $\phi=0$ and $|\psi|=M$. As the
value of $\phi^2$ rises, the minima of the $\psi$ component at fixed
$\phi$ shift inwards until eventually, when $\phi^2$ reaches a critical value
$\phi_{c}^2$, the minima meet at $\psi=0$ and the symmetry is restored. The
value of $\phi_{c}^2$ is
\begin{equation}
\phi_{c}^2 = M^2 \frac{\lambda}{\lambda'} \,.
\end{equation}
For convenience we shall always take positive $\phi$.

The equations of motion for the scalar fields in a flat, isotropic Friedmann
universe are
\begin{eqnarray}
\label{fried}
H^{2} & = & \frac{1}{3} \left( \frac{1}{2} \dot{\phi}^{2} + \frac{1}{2}
\dot{\psi}^{2} + V ( \phi , \psi) \right) \,; \\
\label{phieq}
\ddot{\phi} + 3H \dot{\phi} & = & - \frac{\partial V(\phi,\psi)}{\partial
\phi}\,;\\
\label{psieq}
\ddot{\psi} + 3H \dot{\psi} & = & - \frac{\partial V(\phi,\psi)}{\partial
\psi} \,,
\end{eqnarray}
where a dot indicates a derivative with respect to time. Throughout
this paper units are chosen such that $c = \hbar = m_{{\rm Pl}}^{2}/8 \pi =
1$. These equations are invariant under a simultaneous rescaling
of the potential and the time variable, $V \rightarrow \alpha V$ and $t
\rightarrow t/\sqrt{\alpha}$.

The initial conditions are assumed to be chaotic. We are interested
primarily in the subset of initial conditions where $\psi$ reaches its
minimum at $\psi= 0$ while $\phi$ is still large, and then $\phi$ evolves
down the $\psi = 0$ channel. (Initial conditions where the field falls
directly into the true vacuum give phenomenology no different from single
field models.) With $\psi = 0$ and $\phi^2 > \phi_c^2$, the effective theory
is of a single coupled field $\phi$ in a potential
\begin{equation}
\label{eqpot}
V ( \phi ) = \frac{\lambda}{4} M^{4} + \frac{\lambda''}{4} \phi^{4} \,.
\end{equation}
We shall frequently refer to the first term as the false vacuum energy.
Once $\phi^2$ reaches $\phi_{c}^2$, the $\psi$ field develops degenerate
minima which move out towards the overall minima of the system at
$\phi = 0 , |\psi| = M$. Inflation may end either when the $\psi$ field moves
away  from $\psi=0$ (which requires $\phi^2<\phi_{c}^2$) or when the
potential in the $\phi$ direction becomes too steep to sustain inflation.

Since the equations of motion are invariant under a simultaneous
rescaling of the potential and the time variable, it is useful to write
Eq.~(\ref{eqpot}) in the form
\begin{equation}
V=\frac\lambda4 M^4 (1+B\phi^4) \,,
\end{equation}
where
\begin{equation}
B \equiv \frac{\lambda''}{\lambda M^4} \,.
\end{equation}
The rescaling guarantees that while on the $\psi = 0$ trajectory the dynamics
can depend only on $B$. Ignoring for the moment the instability, this
potential determines the evolution of $\phi$  through Eq.~(\ref{phieq}).
One approach to its solution is given by the usual slow-roll approximation
\begin{equation}
\label{ATT}
3H \dot{\phi} \simeq - V' \,,
\end{equation}
where primes indicate derivatives with respect to $\phi$. Two dimensionless
functions may be defined, whose values indicate the validity or otherwise of
the slow-roll approximation. These functions are
\begin{eqnarray}
\epsilon & \equiv & \frac{1}{2} \left(\frac{V'}{V}\right)^{2}\,;\\
\eta & \equiv & \frac{V''}{V}
 \,,
\end{eqnarray}
where the prime indicates a derivative with respect to $\phi$. A necessary,
but unfortunately not sufficient, condition for the slow-roll approximation
to apply is that
\begin{equation}
\epsilon , |\eta| \ll 1 \,.
\end{equation}

In any case, we shall find that the slow-roll approximation is inadequate for
determining the full range of possibilities this model encompasses.

\section{Interrupted inflation}

A notable feature of the quartic potential is that inflation can end as
$\eta$ and $\epsilon$ rise above unity, and then restart again as they fall
back below unity\footnote{Contrary to the claim of footnote 3 in
\cite{CLLSW}, such double epoch inflation can also occur for a potential of
the form $V_0 + m^2 \phi^2/2$, but there only $\epsilon$ can fall back
below unity. As a result inflation can restart only for a fraction
of a Hubble time in that case.}. Such a possibility was mentioned by Linde
\cite{LIN2SC2}, though he was actually considering a different dynamical
regime to the one we will discuss; in his scenario
the field would exhibit oscillations during which the
instability was ineffective, inflation restarting once the oscillations were
sufficiently damped by the expansion. We shall only consider parameters where
the instability acts rapidly (see Section \ref{INST}), implying that
inflation restarts without an intermediate oscillatory regime.

For our potential, Eq.~(\ref{eqpot}), we have
\begin{equation}
\eta=\frac{12B\phi^2}{1+B\phi^4} \,,
\end{equation}
and
\begin{equation}
\frac{ \epsilon }{ \eta } = \frac{2}{3} \frac{ B\phi^4}{1+B\phi^4}<\frac23
\,.
\end{equation}
Since $\eta$ is always the larger of the two quantities we
shall concentrate on its value. For sufficiently large $\phi$,
$\eta$ will be less than unity. As $\phi$ falls, $\eta$ will initially rise
until it reaches a maximum beyond which it will fall back towards zero.
Provided that
\begin{equation}
\label{Eqgap}
B>\frac{1}{36} \simeq 0.028 \,,
\end{equation}
that maximum will be above unity; the
values of $\phi$ at which $\eta$ equals unity are then given by
\begin{equation}
\phi_{\pm} = \sqrt{6} \left( 1 \pm \sqrt{ 1 - \frac{ 1}{ 36
B}} \right)^{ \frac{1}{2} } \,.
\end{equation}

As $\eta$ initially rises above unity the slow-roll conditions cease
to be satisfied, suggesting inflation may terminate. However, unless the
instability $\phi_c$ lies between $\phi_+$ and $\phi_-$, $\eta$ will
once more fall below unity and the slow-roll conditions shall once more be
satisfied. It is therefore possible that inflation may be able to restart,
though only if the slow-roll `attractor' Eq.~(\ref{ATT}) can be attained in
time. Unfortunately, the slow-roll approximation is necessarily a poor guide
to the evolution under such circumstances \cite{LPB}, and we resort to
numerical simulation.

Numerical simulation indicates instead that {\em ignoring the instability}
there are three regimes of behaviour.
\begin{description}
\item[$0 < B \lesssim 0.19$~:] Inflation proceeds as a single phase, with no
interruption. Although near the top of the range both $\eta$ and $\epsilon$
momentarily exceed unity, they do not do so for long enough for inflation to
terminate.
\item[$0.19 \lesssim B \lesssim 2$~:] Inflation stops at
$\phi\simeq\phi_+\simeq \sqrt{12}$, and then restarts, giving two
separate epochs. As $B$ increases within this range, the value of $\phi$ at
which inflation restarts approaches zero.
\item[$2 \lesssim B$~:]
Inflation stops at
$\phi\simeq\phi_+\simeq \sqrt{12}$ and never  restarts.
\end{description}

Consequently, the regime of double epoch inflation is considerably narrower
than the slow-roll approximation suggests; it needs a larger value of $B$ and
the slow-roll approximation gives no hint that there might be an upper limit
on $B$ beyond which it does not occur, especially one so close to the lower
limit.

These results are important for our later classification of the possible
behaviours when the presence of the instability is taken into account.

\section{Density perturbations}

The most important constraint placed on the parameters of inflationary models
comes from the size of the density perturbations that the models generate.
The particular density perturbations that we are interested in are those
produced about sixty $e$-foldings before the end of inflation, as they can
induce observable microwave background anisotropies. The density contrast at
this time is given by \cite{LL2}
\begin{equation}
\delta_{H}^{2} = \frac{1}{ 300 \pi^{2} } \frac{V_{60}}{\epsilon_{60}} \,,
\end{equation}
where $V_{60}$ is the value of the potential 60 $e$-foldings before the
end of inflation and $\epsilon_{60}$ is the value of $\epsilon$ at the same
time. Usually it is assumed that these $e$-foldings are $e$-foldings in the
scale factor $a$. But in actual fact it is the quantity $aH$ that must
increase by 60 $e$-folds after structure formation scales cross outside the
Hubble radius. As $H$ is approximately a
constant during  inflation it is usually satisfactory to consider just
the increase in $a$, but this need not be so  if inflation is interrupted.

To evaluate $\phi_{60}$ we need the number of $e$-foldings occurring
when inflation proceeds continuously between two values of $\phi$
\begin{equation}
N ( \phi_{1} , \phi_{2} ) \equiv \ln \frac{a_{2}}{a_{1}} \simeq - \int_{
\phi_{1}}^{\phi_{2}} \frac{V}{V'} d \phi \,.
\end{equation}
If inflation proceeds without interruption after $\phi_{60}$ then
\begin{equation}
N ( \phi_{60},\phi_{E} ) = 60 = \frac{1}{8} \left[
B^{-1}(\phi_E^{-2}-\phi_{60}^{-2}) +(\phi_{60}^{2}-\phi_{E}^{2})
\right] \,,
\end{equation}
where $\phi_{E}$ is the value of $\phi$ where inflation ends.
We will consider the effect of an interruption later.

The value of $\delta_{H}$ as given by the recent analysis of the COBE data by
G\'{o}rski {\it et al} \cite{Gor} is
\begin{equation}
\delta_{H} = 2.3 \times 10^{-5} \,.
\end{equation}
This value applies as long as the spectral index $n$ of the density
perturbations is close to unity, and the amplitude of gravitational waves is
sufficiently small. One is also interested in the size of the
spectral index, which in the slow-roll approximation is given by
\begin{equation}
n = 1 + 2 \eta_{60} - 6 \epsilon_{60} \,.
\end{equation}

\section{The instability at $\phi_{c}$}
\label{INST}

Many of the interesting features of this model result from the presence of
the instability in the $\psi$ field which occurs at values of $\phi^2$
below the critical value $\phi_{c}^2$.
If $\phi^2$ is below $\phi_c^2$, the point $\psi=0$ is no longer  stable
and in a wide range of parameter space inflation becomes impossible.
As discussed in Refs.~\cite{LIN2SC,CLLSW,LIN2SC2} for the quadratic case, one
can establish this fact by supposing that  on the contrary inflation does
occur, and demonstrating a contradiction.  We will do the same for the
quartic potential. As with the quadratic case the regime of parameter  space
over which inflation is shown not to occur is large, but not necessarily
optimal; in other words, it might well be possible to extend it by more
detailed arguments. We will consider only the case where the false vacuum
dominates the potential when the instability is encountered, since in the
opposite case one does not expect the false vacuum to significantly affect
the dynamics of inflation.

Suppose that inflation continues for a Hubble time after the instability
is encountered. To  demonstrate that this supposition leads to a
contradiction we need to show that it has the following consequences
\begin{enumerate}
\item The $\psi$ field rolls down to its minimum during this time.
\item Inflation does not continue in the new minimum.
\end{enumerate}

As we are only trying to establish a contradiction we can suppose that the
$\phi$ and $\psi$ fields are both homogeneous, since inhomogeneity could
hardly lead to greater stability. For small values of the field $\psi$, the
$\psi^{3}$ term in Eq.~(\ref{psieq}) may be neglected, and hence the equation
of motion for the $\psi$ field is
\begin{equation}
\ddot{\psi} + 3 H \dot{\psi} + M_{\psi}^{2} (\phi) \psi = 0 \,,
\end{equation}
where
\begin{equation}
M_{\psi}^{2} (\phi) = \lambda' (\phi^{2} - \phi_{c}^{2}) \,.
\end{equation}
More or less independently of its initial value, $\psi$ will  roll down
within a Hubble time if $|M_{\psi}^{2}|$ exceeds $H^{2}$. Using
Eq.~(\ref{ATT}) with vacuum dominated value $H^2=\lambda M^4/12$,
one finds that after one Hubble time
\begin{equation}
\phi^2=\phi_c^2\left(1+\frac{8\lambda''}{\lambda' M^2}
\right)
\,,
\end{equation}
so that
\begin{equation}
\frac{|M_\psi^2|}{H^2}=\frac{96\lambda''}{\lambda'M^4}
\left(1+\frac{8\lambda''}{\lambda'M^2} \right)^{-1} \,.
\end{equation}
Thus $\psi$ will  fall  to its minimum within a Hubble time if
\begin{equation}
96\gg \frac{\lambda'M^4}{\lambda''}+8M^2 \,.
\end{equation}
In a sensible  particle physics model $M\lesssim 1$,  so this constraint is
just
\begin{equation}
96\gg \frac{\lambda'M^4}{\lambda''} \,.
\label{waterfall}
\end{equation}

It remains to show that inflation does not proceed in the new minimum. If it
did, $\phi$ would be slowly varying so the potential in the new minimum would
be approximated by setting $\phi$ equal to a constant. This gives
\begin{eqnarray}
V(\phi) & = & \frac{\lambda'' \phi^{4}}{4} + \frac{\lambda M^{4}}{4} \left(1
- \left( 1 - \frac{\phi^{2}}{\phi_{c}^{2}} \right)^{2} \right) \,;\\
V'(\phi) & = & \lambda'' \phi^{3} + \lambda M^{4} \frac{\phi}{\phi_{c}^{2}}
\left( 1 - \frac{\phi^2}{\phi_{c}^{2}} \right) \,;\\
V''(\phi) & = & 3 \lambda'' \phi^{2} + \frac{\lambda M^{4}}{\phi_{c}^{2}}
\left( 1 - 3 \frac{\phi^{2}}{\phi_{c}^{2}} \right) \,.
\end{eqnarray}
Inflation cannot proceed (for more than a Hubble time or so) if $\eta\gtrsim
1$ or $\epsilon\gtrsim 1$, and assuming the vacuum domination condition
$\lambda''\phi_c^4\ll \lambda M^4$ one verifies that one of these
inequalities is satisfied for all $\phi<\phi_c$  if
\begin{equation}
\phi_c^2\lesssim 8 \,.
\end{equation}
Since  the vacuum domination condition may be  written $\phi_c^4\ll
B^{-1}$,  this  condition is  redundant unless  $B$ is very small.
Ignoring it, our conclusion is that in the vacuum dominated regime
inflation cannot  proceed with $\phi^2<\phi_c^2$, if Eq.~(\ref{waterfall}) is
satisfied. We assume this condition from now on.

It should be emphasised  that the nature of the phase transition that occurs
when $\phi^2$ falls below $\phi_c^2$ is likely to be very complicated, with
both the $\psi$ and $\phi$ fields extremely inhomogeneous \cite{CLLSW}.
There is no reason to suppose that it resembles the orderly  sequence of
events described above (where the slow roll of a homogeneous  $\phi$ field
triggers the fast roll of a homogeneous $\psi$ field), which was
postulated only to show that a contradiction followed.

\section{The Epochs of Inflation}

Let us now consider the possible regimes of behaviour. These are most
effectively classified in accordance with the way inflation ends, with those
classes subdivided according to the particular way inflation leads to that
end. This description is summarised in Table I.

Let us introduce some new notation. The precise location at which inflation
ceases on the $\psi = 0$ trajectory shall be denoted $\phi_{{\rm stop}}$. If
$B \lesssim 0.19$ it does not exist and can be allocated the default value
zero. When it exists, $\phi_+$ provides a good approximation to it. If
inflation can restart further down the potential, the value at which this
happens is denoted $\phi_{{\rm restart}}$. Usually, $\phi_-$ does {\em not}
provide a good approximation due to the failure of the inflationary
attractor. As $B$ approaches 2, it tends to zero.
\begin{description}
\item[A:] Inflation ends by steepening of the potential\\
In this regime, inflation terminates at $\phi_{{\rm stop}}$ and does not
restart. This implies $B \gtrsim 0.19$. The instability must be below
$\phi_{{\rm stop}}$, and if $B \lesssim 2$, giving the regime where inflation
can potentially restart, the instability must be at $\phi^2 > \phi_{{\rm
restart}}^2$.
\item[B:] Inflation ends by instability
\begin{enumerate}
\item {\it Single epoch inflation:}
If $B \lesssim 0.19$; the potential is flat enough for inflation to occur
everywhere on it and instability is the only way inflation can end.
If $B \gtrsim 0.19$, then one still has only a single epoch of inflation if
the instability is positioned above $\phi_{{\rm stop}}$.
\item {\it Double epoch inflation:} With $0.19 \lesssim B \lesssim 2$,
inflation can restart provided the instability is located at small enough
$\phi$. Phenomenologically, this is further divided by whether the scales of
astrophysical interest, sixty $e$-foldings from the end of inflation, crossed
outside the Hubble radius during the first or the second period of
inflation, or both.
\end{enumerate}
\end{description}

\subsection{Inflation ends by steepening of the potential}

The condition that the potential can steepen enough to terminate inflation is
sufficient to guarantee that the false vacuum term is small even right to the
end of inflation, being at most four percent of the quartic term at
$\phi_+$. Apart from minor corrections when $B$ gets close to its limiting
value, the situation is therefore identical to the original $\phi^{4}$ model
of chaotic inflation.  To calculate the density perturbations we need to
calculate $\phi_{60}$, the value of $\phi$ sixty $e$-foldings before the end
of inflation, which is given in this limit by
\begin{equation}
\phi_{60} = \sqrt{ 480 + \phi_{+}^{2} } = \sqrt{492} \,.
\end{equation}
Hence the COBE constraint gives
\begin{equation}
\delta_{H}^{2} = \frac{1}{300 \pi^{2}} \frac{ (492)^{3} \lambda''}{32} \,,
\end{equation}
which determines the value of $\lambda''$ as
\begin{equation}
\label{standard}
\lambda'' = 4.2 \times 10^{-13} \,.
\end{equation}
This is the standard result, needing no further comment.

\subsection{Inflation ends by instability}

\subsubsection{Single epoch inflation}

For $B \lesssim 0.19$ inflation can occur everywhere on the potential, and
given
the freedom to place the instability anywhere on it a variety of scenarios
result. We shall concentrate on the two limiting cases, where the last sixty
$e$-foldings of inflation are dominated by either the quartic term or the
false vacuum term.

If $B\gtrsim 0.19$ then we
need $\phi_c^2>\phi_+^2\simeq 12$. This guarantees
domination by the quartic term because during inflation
$B\phi^4>B\phi_c^4>144 B \gg1$.

With domination of the quartic term, the evolution of the system is not
significantly affected by the occurrence of the instability, since the
decay of the false vacuum term adds only small corrections to the
dominant quartic coupling term. Hence this case is identical to the
conventional chaotic inflation model and requires no further discussion.

Domination by the false vacuum term, which requires $B\lesssim 0.19$, is
more interesting, and corresponds to the condition
\begin{equation}
\phi^4_{60}\ll B^{-1} \,.
\end{equation}
Under these circumstances,
\begin{equation}
\phi_{60}=\phi_c\left( 1-480 B \phi_c^2 \right)^{-1/2} \,,
\end{equation}
so false vacuum domination implies
\begin{equation}
480B\phi_c^2 \ll \frac{480B^{1/2}}{1+ 480 B^{1/2}} < 1 \,.
\end{equation}
Hence $\phi_{60}\simeq \phi_c$. This condition may be written
\begin{equation}
\frac{480\lambda''}{\lambda'M^2}\ll 1 \,,
\label{480}
\end{equation}
and the vacuum domination condition becomes
\begin{equation}
\lambda\lambda''\ll \lambda'^2 \,.
\end{equation}

The density perturbation formula gives
\begin{equation}
\delta_{H}^{2} = \frac{1}{32 \times 300 \pi^{2}} \frac{\lambda'^{3} M^{6}}{
\lambda''^{2}}  \,,
\end{equation}
which using COBE places the constraint on the parameters that
\begin{equation}
\lambda' M^{2} = 0.037 \lambda''^{\frac{2}{3}}  \,.
\end{equation}
Combining this with Eq.~(\ref{480}) gives
\begin{equation}
\lambda''\ll 10^{-12} \,.
\end{equation}
Hence this regime does not evade the usual smallness required of the quartic
coupling.

In this regime $\epsilon_{60} \ll \eta_{60}$, so the spectral index is
greater than unity. However, the vacuum domination constraint prevents its
absolute value being significantly higher than unity. A more detailed
investigation along the lines of Ref.~\cite{CLLSW} would be needed to examine
the behaviour as one leaves perfect false vacuum domination.

\subsubsection{Double epoch inflation}

In this rather limited region of parameter space, inflation occurs in two
separate epochs between which the potential is temporarily too steep to
sustain inflation. Its phenomenology depends crucially on the history of the
scales of astrophysical interest. There are three possibilities, any of
which can be realised given suitable adjustment of $\phi_c$, except near the
largest values of $B$ where the second epoch may not be able to support many
$e$-foldings between the restart and $\phi = 0$.

The simplest possibility is that the entire last sixty or so $e$-foldings
occur during the second epoch of inflation; this makes the first epoch
astrophysically irrelevant and is extremely similar to the single epoch
inflation model discussed above, needing no further discussion here.

The most complex is where at least some of the interesting scales cross
outside the Hubble radius near the end of the first epoch, {\em re-enter}
during the interval between inflationary epochs and then exit again during
the second epoch. Such a combination greatly complicates the calculation of
the density perturbations, because the first Hubble radius crossing will
inevitably populate some of the modes, invalidating the usual vacuum
hypothesis which is used as an initial condition for the standard
calculation. As a result, the spectrum produced may have an unusual
amplitude, or scale dependence, or be nongaussian. As considerable tuning is
required to make the interesting scales coincide with the suspension of
inflation, and
because of the complexity of the scenario, we shall not address it further.

The final option is that the scales of interest crossed outside the Hubble
radius for the first and only time during the first epoch of inflation.

It is important to note before progressing any further that when one talks
about $e$-foldings, it is usually implicitly assumed that $H$ is constant.
In general, the number of $e$-foldings is not really a condition on the scale
factor $a(t)$, but actually a condition on the quantity $a(t)H(t)$. Certainly
during the suspension of inflation it is important to take this into account;
a general formalism to do this is discussed in \cite{LPB} though we shall not
require it here.

With $B$ in the required range, numerical simulations show that the actual
reduction of $aH$ during the intermediate regime is always small, at most
around one $e$-folding. Consequently this aspect of the intermediate regime
can be more or less ignored. [We note that for the most complex scenario
mentioned above that this implies that only some of the scales of
astrophysical interest can cross back inside during the intermediate era,
emphasising that the results there will be far from standard.] Operationally,
then, all we need do is assume that $60 - N_1$ $e$-foldings occur in the
second epoch, where $N_1$ would have to be determined numerically taking into
account the intermediate era. Then $\phi_{60}$ is given by
\begin{equation}
\phi_{60} = \sqrt{\phi_{+}^{2} + 8 N_{1}} = \sqrt{12 + 8 N_{1}} \,.
\end{equation}
Hence COBE gives for the value of $\lambda''$
\begin{equation}
\label{straddle}
\lambda'' = \frac{5 \times 10^{-5}}{(12+8N_{1})^{3}} \,.
\end{equation}
Taking $N_1 = 60$ recovers Eq.~(\ref{standard}). In this regime the
self-coupling will not be as small as in the standard single field case,
though it will still be small in an absolute sense.

One must be careful though not to push this too far, because bringing
$\phi_{60}$ closer to $\phi_+$ reduces the spectral index, which is given by
\begin{equation}
n = 1 - 3/N_1 \,.
\end{equation}
As gravitational waves are also produced, a conservative assessment of the
observations \cite{LLPARIS} suggests $n$ must exceed 0.8 and hence $N_1$
should exceed 15.

An interesting effect of the intermediate regime is that $H$ will be
reduced during it. This offers an opportunity for $V_{{\rm end}}$ to be much
lower than $V_{60}$; if inflation is continuous then this cannot usually
be achieved without unacceptably distorting the density perturbation
spectrum. The hybrid model provides a justifiable way of inserting a feature
into the potential to do this. If recent suggestions \cite{REHEAT} that
reheating may be considerably more efficient than previously thought are
borne out, and given that for most models COBE implies $V_{60}$ is in the
vicinity of $10^{16}$ GeV, such a possibility may be one way of evading the
gravitino bound \cite{GRAVITINO} which imposes an upper limit on the reheating
temperature in supersymmetric theories\footnote{Another is the false vacuum
regime, where although $V_{{\rm end}}$ and $V_{60}$ are very close, they are
both very small.}.

\section{Conclusions}

We have studied the inflationary model defined by Eq.~(2) in detail.
During inflation
the potential is proportional to $1+B\phi^4$, with the constant term arising
because a non-inflaton field $\psi$ sits in a false vacuum. When $\phi$
falls below a critical value the vacuum is destabilised, and if it dominates
inflation then ends as the fields adjust to their true vacuum values.

We have found that the delineation of the parameter space with respect to the
various ways in which inflation can proceed is unexpectedly complex.
In particular, there is the possibility that inflation can be interrupted and
then restart, which does not occur in any model that has been studied before
(discounting restart for much less than a Hubble time). What happens is that
inflation ends first by the usual fast roll mechanism,  and then starts again
to end finally when the vacuum destabilises. We have investigated this
possibility in some detail, first qualitatively using the usual slow-roll
conditions, and then quantitatively by evolving the inflaton
field  numerically. Numerical evolution of this kind has been required very
seldom
in the past, but we have found it to be essential for this model.

An intriguing possibility, which we have noted but not
explored, is that some of the cosmological scales of interest might leave
the horizon during the first epoch of inflation, re-enter it
and then leave it again during the second epoch. If that happened
the first  epoch would create inflaton particles, so that the
final adiabatic density perturbation associated with the inflaton field
fluctuation would be quite  different from the usual form.

\section*{Acknowledgements}

DR is supported by PPARC (UK) and ARL by the Royal Society. We acknowledge
useful conversations with Ed Copeland, Ewan Stewart and David Wands. ARL
acknowledges the use of the Starlink computer system at the University of
Sussex.

\newpage
\begin{table}
\centering
\caption[Table]{The different dynamical regimes are indexed in accordance
with the subsection in which they are discussed. Those regions marked as
having no regime correspond to inaccessible parameter choices; in the first
row $\phi_{{\rm stop}}$ doesn't exist and has the notional value zero, while
in the last row $\phi_{{\rm restart}}$ does not exist.}
\begin{tabular}{|l||c|c|c|}
& $\phi_c^2 > \phi_{{\rm stop}}^2$ & $\phi_{{\rm stop}}^2 > \phi_c^2 >
\phi_{{\rm restart}}^2$ & $\phi_{{\rm restart}}^2 > \phi_c^2$ \\
\hline
\hline
$B \lesssim 0.19$ & {\bf B1} & No regime & No regime \\
\hline
$0.19 \lesssim B \lesssim 2$ & {\bf B1} & {\bf A} & {\bf B2} \\
\hline
$2 \lesssim B$ & {\bf B1} & {\bf A} & No regime \\
\end{tabular}
\end{table}


\begin{references}
\bibitem{CHAOTIC} A. D. Linde, Phys. Lett. {\bf B129}, 177 (1983);
	A. D. Linde, {\em Particle Physics and Inflationary Cosmology},
	Harwood Academic, Chur, Switzerland (1990).
\bibitem{LaSteinhardt} D. La and P. J. Steinhardt, Phys. Rev. Lett. {\bf 62},
	376 (1989).
\bibitem{extinf} P. J. Steinhardt and F. S. Accetta, Phys. Rev. Lett. {\bf
	64}, 2740 (1990); J. D. Barrow and K. Maeda, Nucl. Phys. {\bf B341},
	294 (1990); A. R. Liddle and D. Wands, Phys. Rev. D{\bf 45}, 2665
	(1992); A. M. Laycock and A. R. Liddle, Phys. Rev. D{\bf 49},
	1827 (1994).
\bibitem{Linde90} A. Linde, Phys. Lett. {\bf B249}, 18 (1990).
\bibitem{AdamsFreese} F. C. Adams and K. Freese, Phys. Rev. D{\bf 43}, 353
	(1991).
\bibitem{LIN2SC} A. D. Linde, Phys. Lett. {\bf B259}, 38 (1991).
\bibitem{LL2} A. R. Liddle and D. H. Lyth, Phys. Rep. {\bf 231}, 1 (1993).
\bibitem{LIN2SC2} A. D. Linde, Phys. Rev. D{\bf 49}, 748 (1994).
\bibitem{MML} S. Mollerach, S. Matarrese and F. Lucchin, Phys. Rev. D{\bf
	50}, 4835 (1994).
\bibitem{CLLSW} E. J. Copeland, A. R. Liddle, D. H. Lyth, E. D. Stewart
	and D. Wands, Phys. Rev. D{\bf 49}, 6410  (1994).
\bibitem{ewan} E. D. Stewart, ``Inflation, Supergravity and Superstrings'',
	to appear, Phys. Rev. D (1994), hep-ph/9405389; E. D. Stewart,
	``Mutated Hybrid Inflation'', Kyoto preprint (1994),
	astro-ph/9407040; D. H. Lyth and E. D. Stewart, ``False Vacuum
	Chaotic Inflation: The New Paradigm?'', to appear, proceedings of
	`Birth of the Universe and Fundamental Physics', Ed.
	F. Occhionero (1994), hep-ph/9408324; E. D. Stewart, ``A General
	Supergravity Formalism for a Naturally Flat Inflaton Potential'',
	to appear, proceedings of the 7th Marcel Grossman meeting (1994),
	hep-ph/9408302.
\bibitem{WANG} Y. Wang, ``Polynomial Hybrid Inflation'', Fermilab preprint
	FERMILAB-Pub-94/086-A (1994), astro-ph/9405044.
\bibitem{LPB} A. R. Liddle, P. Parsons and J. D. Barrow, ``Formalizing the
	Slow-roll Approximation in Inflation'', to appear, Phys. Rev. D
	(1994), astro-ph/9408015.
\bibitem{Gor} K. G\'{o}rski {\it et al}, Astrophys. J. Lett. 430, L89 (1994).
\bibitem{LLPARIS} D. H. Lyth and A. R. Liddle, to appear in Proceedings of
	Journ\'{e}e Cosmologie, Paris 1994 (World Scientific, Singapore).
\bibitem{REHEAT} L. Kofman, A. D. Linde and A. A. Starobinsky, to appear,
	Phys. Rev. Lett. (1994), hep-th/9405187; Y. Shatanov, J. Traschen
	and R. Brandenberger, ``Universe Reheating after Inflation'',
	Brown preprint (1994), hep-ph/9407247.
\bibitem{GRAVITINO} J. Ellis, D. V. Nanopoulos and S. Sarkar, Nucl. Phys.
	{\bf B259}, 175 (1985); M. Kawasaki and K. Sato, Phys. Lett.
	{\bf B189}, 23 (1987); M. H. Reno and D. Seckel, Phys. Rev. D{\bf
	37}, 3441 (1988); J. Ellis {\it et al}, Nucl. Phys. {\bf B373},
	399 (1992).
\end{references}
\end{document}